\documentclass[conference,letterpaper]{IEEEtran}

\usepackage{graphicx,epsf,psfrag}
\usepackage{amsmath,amssymb}
\usepackage{amsfonts}
\usepackage{mathrsfs}
\usepackage{etoolbox}
\usepackage[noadjust]{cite}
\usepackage{comment}
\usepackage{multirow}

\usepackage{latexsym}

\newcommand{\beq}{\begin{equation}}
\newcommand{\eeq}{\end{equation}}
\newcommand{\be}{\begin{equation}}
\newcommand{\ee}{\end{equation}}
\newcommand{\eps}{\epsilon}

\newcommand{\bi}{\begin{itemize}}
\newcommand{\ei}{\end{itemize}}

\newcommand{\calA}{\mathcal{A}}

\newcommand{\calE}{\mathcal{E}}

\newcommand{\calL}{\mathcal{L}}

\newcommand{\calS}{\mathcal{S}}

\newcommand{\bbN}{\mathbb{N}}

\newcommand{\bbP}{\mathbb{P}}

\newcommand{\bbR}{\mathbb{R}}

\DeclareMathAlphabet{\mathbsf}{OT1}{cmss}{bx}{n}
\DeclareMathAlphabet{\mathssf}{OT1}{cmss}{m}{sl}

\DeclareSymbolFont{bsfletters}{OT1}{cmss}{bx}{n}  
\DeclareSymbolFont{ssfletters}{OT1}{cmss}{m}{n}
\DeclareMathSymbol{\bsfGamma}{0}{bsfletters}{'000}
\DeclareMathSymbol{\ssfGamma}{0}{ssfletters}{'000}
\DeclareMathSymbol{\bsfDelta}{0}{bsfletters}{'001}
\DeclareMathSymbol{\ssfDelta}{0}{ssfletters}{'001}
\DeclareMathSymbol{\bsfTheta}{0}{bsfletters}{'002}
\DeclareMathSymbol{\ssfTheta}{0}{ssfletters}{'002}
\DeclareMathSymbol{\bsfLambda}{0}{bsfletters}{'003}
\DeclareMathSymbol{\ssfLambda}{0}{ssfletters}{'003}
\DeclareMathSymbol{\bsfXi}{0}{bsfletters}{'004}
\DeclareMathSymbol{\ssfXi}{0}{ssfletters}{'004}
\DeclareMathSymbol{\bsfPi}{0}{bsfletters}{'005}
\DeclareMathSymbol{\ssfPi}{0}{ssfletters}{'005}
\DeclareMathSymbol{\bsfSigma}{0}{bsfletters}{'006}
\DeclareMathSymbol{\ssfSigma}{0}{ssfletters}{'006}
\DeclareMathSymbol{\bsfUpsilon}{0}{bsfletters}{'007}
\DeclareMathSymbol{\ssfUpsilon}{0}{ssfletters}{'007}
\DeclareMathSymbol{\bsfPhi}{0}{bsfletters}{'010}
\DeclareMathSymbol{\ssfPhi}{0}{ssfletters}{'010}
\DeclareMathSymbol{\bsfPsi}{0}{bsfletters}{'011}
\DeclareMathSymbol{\ssfPsi}{0}{ssfletters}{'011}
\DeclareMathSymbol{\bsfOmega}{0}{bsfletters}{'012}
\DeclareMathSymbol{\ssfOmega}{0}{ssfletters}{'012}

\newcommand{\hatm}{\hat{m}}

\newcommand{\barR}{\bar{R}}
\newcommand{\barS}{\bar{S}}
\newcommand{\barT}{\bar{T}}

\newcommand{\barV}{\bar{V}}

\newcommand{\barZ}{\bar{Z}}

\newcommand{\floor}[1]{\lfloor{#1}\rfloor}

\DeclareMathOperator*{\minimize}{minimize}

\DeclareMathOperator{\tr}{tr}

\ifcsmacro{theorem}{}{
\newtheorem{theorem}{Theorem}
\newtheorem{lemma}{Lemma}

}

\newcommand{\qednew}{\nobreak \ifvmode \relax \else
      \ifdim\lastskip<1.5em \hskip-\lastskip
      \hskip1.5em plus0em minus0.5em \fi \nobreak
      \vrule height0.75em width0.5em depth0.25em\fi}

\usepackage{color}
\usepackage{hyperref}

\usepackage{bbm}

\allowdisplaybreaks

\title{Nobody Expects a Differential Equation:\\
Minimum Energy-Per-Bit for the Gaussian Relay Channel with Rank-1 Linear Relaying}

\author{%
  \IEEEauthorblockN{Oliver Kosut}
  \IEEEauthorblockA{School of Electrical, Computer \\and Energy Engineering \\
                    Arizona State University\\
                    Tempe, AZ, USA\\
                    Email: okosut@asu.edu}
  \and
  \IEEEauthorblockN{Michelle Effros}
  \IEEEauthorblockA{Department of Electrical Engineering\\ 
                    California Institute of Technology\\
                    Pasadena, CA, USA\\
                    Email: effros@caltech.edu}
    \and
    \IEEEauthorblockN{Michael Langberg}
    \IEEEauthorblockA{Department of Electrical Engineering \\
            University at Buffalo \\
            Buffalo, NY, USA\\
            email: mikel@buffalo.edu}
}

\begin{document}

\maketitle

\vspace{-.4in}

\begin{abstract}
Motivated by the design of low-complexity low-power coding solutions for the Gaussian relay channel, this work presents an upper bound on the minimum energy-per-bit achievable on the Gaussian relay
channel using rank-1 linear relaying.
Our study addresses high-dimensional relay codes and presents bounds that outperform prior known bounds using 2-dimensional schemes.
A novelty of our analysis ties the optimization problem at hand to the solution of a certain differential equation which, in turn, leads to a low energy-per-bit achievable scheme.
\end{abstract}

\section{Introduction}
As communication capabilities
are built into an increasingly diverse
array of technologies,
coding strategies are needed
for more (and often more constrained) communication scenarios.
In some cases, the constraints are not new.
Worries about power limitations, for example,
and characterizations of the minimal energy required, on average,
to reliably deliver each bit of information
across a noisy channel
(the so called ``minimum energy-per-bit'')
date all the way back to Shannon~\cite{Shannon:49}.
Given the proliferation of low-power wireless devices 
with built-in communication capabilities, 
such characterizations are, arguably,
even more relevant today than they were in 1949
when Shannon proposed the question
and derived an asymptotic solution
of the minimum energy-per-bit
for a point-to-point channel
with additive, white Gaussian noise.
Wireless communication devices
motivate much of the ongoing work
on this problem.
Examples of results in this area include, among many others,
characterizations of the minimum energy-per-bit
in the finite blocklength domain
with and without feedback~\cite{PolyanskiyP:11},
in broadcast and interference channels
with correlated information~\cite{Host-Madsen:13},
and in relay channels~\cite{boundsElGamal2006}.

Motivated by wireless communication scenarios
where power constraints are especially restrictive,
this work provides a bound on the minimum energy-per-bit
for the relay channel under a constrained family of codes.
We choose the relay channel
since it is well matched to low-power devices
that may be unable to meet
their communication goals
without the aid of a helper, here described as a relay node.
We maintain low coding complexity by using a constrained family of codes --- namely, where the relay performs only linear operations, and where the transmitter sends a Gaussian vector whose covariance matrix has rank 1.
Our desire to limit coding complexity
is motivated by the observation
that devices with tight power constraints
are often also extremely constrained in computational resources.
In some cases, limits on computation
can be traced back to power constraints;
in particular, while the energy costs
for using a code are typically far smaller than
those for transmission,
at extremely low power
the two can become comparable,
leaving the code operation
to compete for resources with the cost of transmission.\footnote{The
costs of code operation are device specific
and are not included
in any of the prior
minimum energy-per-bit characterizations
cited above or in our characterization.}
Complexity constraints can also arise
when constraints on price, weight, or size
result in limitations
on the computational resources
on board a given communication device.

The main result of this work is an upper bound on
the minimum energy-per-bit
achievable on the Gaussian relay channel
using the family of low-complexity
rank-1 linear relay codes,
described in Section~\ref{sec:problem}. The minimum energy-per-bit achievable by this family of codes can be posed as a non-convex optimization problem. We analyze the necessary optimality conditions for this problem and find that taking a limit of these optimality conditions leads to a system of differential equations. Solutions to this system gives rise to our achievable bound, presented in Section~\ref{sec:main} and proved via a number of lemmas in Sections~\ref{sec:optimality} and~\ref{sec:DE}.

\section{Problem Description}\label{sec:problem}

\emph{Notation:} We use $\log$ to denote log base-2, and $\ln$ for natural log. For any $p\ge 1$, $\|x\|_p$ denotes the $p$-norm of $x$; when the subscript is left off we mean $2$-norm. The set of real numbers is denoted $\bbR$ and the set of positive integers $\bbN$. For an integer $M$, $[M]$ denotes $\{1,2,\ldots,M\}$.

The Gaussian relay channel has two gain parameters $a$ and $b$. At time instance $i$, the transmitted signals at the transmitter and the relay are denoted $X_{i}$ and $X_{ri}$ respectively. The received signals at the relay and receiver are
\be\label{channel_description}
Y_{ri}=aX_i+Z_{ri},\qquad Y_i=X_i+bX_{ri}+Z_{i}
\ee
where $\{Z_i\}$ and $\{Z_{ri}\}$ are independent white Gaussian noise sequences with mean $0$ and variance $1$. We assume without loss of generality that $a$ and $b$ are non-negative. In addition, we assume that both are strictly positive, since if either is $0$, the relay is effectively removed from the network.

An $(n,M)$-code with strict causality at the relay is given by encoding functions at the transmitter and relay
\be
    x^n:[M]\to \bbR^n,\qquad
    x_{ri}:\bbR^{i-1}\to\bbR,\ i=1,\ldots,n
\ee
and a decoding function
\be
\hat{m}:\bbR^n\to[M].
\ee
The probability of error is 
\begin{multline}
P_e^{(n)}=\frac{1}{M}\sum_{m=1}^M\bbP\{\hatm(Y^n)\ne m|X^n=x^n(m),
\\X_{ri}=x_{ri}(Y_r^{i-1})\}
\end{multline}
where $Y_{ri},Y_i$ are related to $X_i,X_{ri}$ by \eqref{channel_description}. The energy-per-bit is denoted
\be
\calE^{(n)}=\frac{\max_{m}\|x^n(m)\|^2+\sup_{y_r^n}\sum_{i=1}^n x_{ri}(y_r^{i-1})^2}{\log M}.
\ee
We say energy-per-bit $\calE$ is \emph{achievable} if there exists a sequence of codes where $P_e^{(n)}\to 0$ and $\limsup \calE^{(n)}\le \calE$. The \emph{minimum energy-per-bit} $\calE^*$ is the infimum of all achievable energies-per-bit.

A linear relay code is one where, at every time instant, the relay sends a linear combination of its received signals. This relaying can be described by a strictly lower-triangular matrix $D$, to ensure that coding at the relay is strictly causal. Under this assumption, it is optimal that the transmitter send a Gaussian vector. Thus, the code can be described by the matrix $D$ and the covariance matrix $\Sigma$ for the transmitted vector. Let $\calE^*_{\text{LR}}$ be the minimum energy-per-bit achievable by linear codes. From results in \cite{boundsElGamal2006}, it follows that $\calE^*_{\text{LR}}$ is equal to
\be\label{ELR}
\inf_{\substack{k\in\bbN,\\ \Sigma\in\calS_+^k,\\ D\in\calL^k}}
\frac{\tr(\Sigma+a^2D\Sigma D^T+DD^T)}{\displaystyle\frac{1}{2}\log\frac{\det((I+abD)\Sigma (I+abD^T)+I+b^2DD^T)}{\det(I+b^2DD^T)}}
\ee
where $\calS_+^k$ is the set of positive semi-definite $k\times k$ matrices, and $\calL^k$ is the set of strictly lower triangular $k\times k$ matrices. In addition, let $\calE^*_{\text{1LR}}$ be the value of \eqref{ELR} where $\Sigma$ is constrained to be rank-1. Obviously $\calE^*\le \calE^*_{\text{LR}}\le\calE^*_{\text{1LR}}$. Our main result will be an upper bound on $\calE^*_{\text{1LR}}$.

\subsection{Prior Bounds}\label{sec:prior_bounds}

In \cite{boundsElGamal2006} it is shown that the block-Markov achievable scheme (i.e., partial decode-forward) from \cite{CapacityCover1979} leads to the upper bound on the minimum energy-per-bit
\be\label{block_Markov}
\calE^*\le 2\ln 2\, \min\left\{1,\,\frac{a^2+b^2}{a^2(1+b^2)}\right\}.
\ee
It is also shown in \cite{boundsElGamal2006} that linear relaying can outperform this bound, using a $k=2$-dimensional linear scheme given by
\be\label{2x2_scheme}
\Sigma=2P_1\left[\begin{array}{cc}\beta & \sqrt{\beta(1-\beta)} \\ \sqrt{\beta(1-\beta)} & 1-\beta\end{array}\right],
\quad
D=\left[\begin{array}{cc}0 & 0\\ d & 0\end{array}\right]
\ee
where $\beta\in[0,1]$ and 
$
d=\sqrt{\frac{2P_2}{2a^2\beta P_1+1}},
$
and where $P_1,P_2$ are the powers used at the transmitter and relay respectively. Note that $\Sigma$ in \eqref{2x2_scheme} has rank 1.

In addition, \cite{boundsElGamal2006} found that the cut-set bound from \cite{CapacityCover1979} leads to the lower bound on the minimum energy-per-bit given by
\be\label{cutset_bound}
\calE^*\ge 2\ln 2\,\frac{1+a^2+b^2}{(1+a^2)(1+b^2)}.
\ee

In \cite{JointKim2012}, an LTI filtering approach was taken to coding for the relay channel. These also constitute linear codes, although for the strictly causal case for channels with constant gains, the achievability bounds do not appear to be any better than those from \cite{boundsElGamal2006}. Linear codes for the Gaussian relay channel were also considered in \cite{JointGohary2013}, which derived methods to simplify the optimization problem for $\Sigma$ and $D$ for a given dimension $k$. This approach is interestingly complementary to ours, in that \cite{JointGohary2013} finds optimal $\Sigma$ for a given $D$, whereas we focus on finding optimal $D$ for a given $\Sigma$. Even with these simplifications, finding optimal $(\Sigma,D)$ pairs is challenging, particularly for large $k$, so we have chosen to omit comparisons in this paper.

\section{Main Result}\label{sec:main}

To state our result, we first need the following lemma, proved in Appendix~\ref{appendix:A0c1}.

\begin{lemma}\label{lemma:A0c1}
Fix any $A_f,B_f>0$ where $A_f/B_f\le a^2$. Let\footnote{Note that $f$ implicitly depends on $A_f,B_f$ through $\phi$.}
\begin{align}
\phi&=A_f B_f+\frac{1}{A_f}-\frac{1}{B_f},\label{c2_def}\\
f(w)&=\frac{\phi w-1+\sqrt{(\phi w-1)^2+4 w^3}}{2w^2}.\label{f_def}
\end{align}
There exists a unique pair $(A_0,\psi)$ such that $A_0\ge A_f$, $\psi>0$ and
\begin{align}
\int_{A_f}^{A_0} \frac{f(w)\,dw}{1+wf(w)^2}&=\frac{A_0}{a \psi}-\frac{1}{B_f},\label{first_integral_eqn}\\
\int_{A_f}^{A_0}\frac{f(w)^2\,dw}{1+wf(w)^2}&=\ln\left(\frac{A_0^3 B_f}{a^4 \psi^2}\right).\label{second_integral_eqn}
\end{align}
\end{lemma}

Given any $A_f,B_f>0$ satisfying $A_f/B_f\le a^2$, define $\calE(A_f,B_f)$ as follows. First let $(A_0,\psi)$ be the pair from Lemma~\ref{lemma:A0c1}. Also let $f(w)$ be the function in \eqref{f_def}. Then define
\be
\calE(A_f,B_f)=\frac{Q_1+Q_2}{\frac{1}{2}\log\frac{A_0}{a^2}\left(\frac{1}{B_f}+A_0B_0-A_fB_f\right)}
\ee
where
\begin{align}
    B_0&=f(A_0),\\
    Q_1&=-\frac{1}{a^2}+\frac{A_0^3 B_f}{a^6 \psi^2},\label{Q1_thm}\\
    Q_2&=-\frac{1}{b^2}+\frac{A_0^3}{a^5 b^2 \psi^3}+\frac{A_0^2(A_fB_f^2-1)}{a^4 b^2 \psi^2 B_f}.\label{Q2_thm}
\end{align}

\begin{theorem}\label{thm:main}
For any $a,b >  0$,
\be
\calE^*_{\text{1LR}}\le\inf_{A_f,B_f>0:A_f/B_f\le a^2}\calE(A_f,B_f).
\ee
\end{theorem}

Theorem~\ref{thm:main} is derived from the optimality conditions for the optimization problem \ref{ELR} when $\Sigma$ is rank-1, then taking a limit as the energy per symbol goes to 0, and analyzing the resulting differential equation (see the below sections for the details). For this reason, we suspect that the bound in Thm.~\ref{thm:main} is optimal or close t optimal among rank-1 linear codes.

\begin{figure}
    \includegraphics[width=\columnwidth]{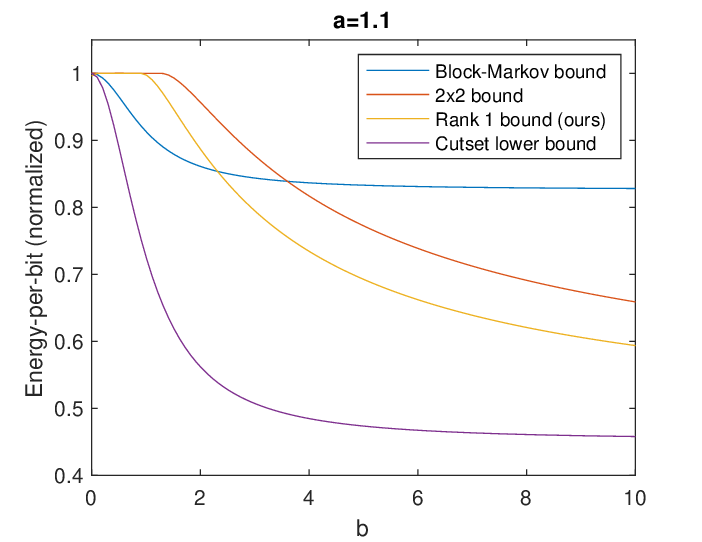}
    \caption{Comparison between bounds on the minimum energy-per-bit for the Gaussian relay channel. Bounds are computed for channel parameters $a=1.1$ and $b\in[0,10]$. The normalized energy-per-bit $\calE/(2\ln 2)$ is plotted for each bound. The block-Markov achievable bound is \eqref{block_Markov}, the $2\times 2$ bound is from \eqref{2x2_scheme}, the rank 1 bound is from Thm.~\ref{thm:main}, and the cut-set bound is \eqref{cutset_bound}.}
    \label{fig:comparison}
\end{figure}

Numerical results for the bound from Thm.~\ref{thm:main} compared to  bounds mentioned in Sec.~\ref{sec:prior_bounds} are shown in Fig.~\ref{fig:comparison}. As shown in the figure, our bound consistently outperforms the $k=2$ bound in \eqref{2x2_scheme}, and it sometimes outperforms the block-Markov bound in \eqref{block_Markov}.

\section{Transforming the Optimization Problem}\label{sec:optimality}

The first step toward proving Thm.~\ref{thm:main} is to transform the optimization problem in \eqref{ELR} under the assumption that $\Sigma$ has rank 1. 
Specifically let $\Sigma=ss^T$ where $s\in\bbR^k$. Let the matrix $G$ be such that $GG^T=(I+b^2DD^T)^{-1}$, and let $v=G^T(I+abD)s$. Then the quantity inside the log in \eqref{ELR} is
\begin{align}
    &\frac{\det\left((I+abD)ss^T(I+abD^T)+I+b^2DD^T\right)}{\det\left(I+b^2DD^T\right)}
\\&=\det \left[G^T\left((I+abD)ss^T(I+abD^T)+I+b^2DD^T\right)G\right]
\\&=\det \left[G^T(I+abD)ss^T(I+abD^T)G+I\right]
\\&=\det(I+vv^T)
\\&=1+\|v\|^2\label{rank1_determinant}
\\&=1+s^T(I+abD^T)GG^T(I+ab)s
\\&=1+s^T(I+abD^T)(I+b^2DD^T)^{-1}(I+abD)s\label{inside_log}
\end{align}
where \eqref{rank1_determinant} follows because the eigenvalues of $I+vv^T$ are $1+\|v\|^2$ and $1$, the latter with $k-1$ repetitions.
We thus define, for any $k$, $s\in\bbR^k$, $D\in\bbR^{k\times k}$,
\begin{multline}\label{EsD_formula}
\calE_{\text{1LR}}(s,D)=\\
\frac{\|s\|^2+a^2\|Ds\|^2+\tr(DD^T)}{\frac{1}{2}\log(1+s^T(I+abD^T)(I+b^2DD^T)^{-1}(I+abD)s)}.
\end{multline}
Also define
\be\label{D_opt}
\calE_{\text{1LR}}^*(s)=\sup_{D\in\calL^k}\  \calE_{\text{1LR}}(s,D).
\ee
Thus
\be
\calE^*_{\text{1LR}}=\sup_{k\in\bbN}\  \sup_{s\in\bbR^k}\  \calE_{\text{1LR}}^*(s).
\ee

With some hindsight, we make the following definitions, which will allow us to rewrite and analyze the optimization problem over $D$ in \eqref{D_opt}:
\begin{align}
u&=Ds,\\
z&=(I+b^2DD^T)^{-1}(s+abu),\\
r&=abs-b^2D^Tz.
\end{align}
Given $s$ and $D$, $z$ is well-defined since $I+b^2DD^T$ is positive definite, and so it is invertible. With these definitions, \eqref{inside_log} becomes $1+(s+abu)^Tz$. In addition,
\begin{align}
s+abu&=(I+b^2DD^T)z=z+D(b^2D^Tz)
\\&=z+D(abs-r)
=z+abu-Dr.
\end{align}
Thus, $Dr=z-s$. Now for any $s$,
 $\calE_{\text{1LR}}^*(s)$ is the optimal value of the optimization problem
\be\label{transformed_optimization}
\begin{array}{ll}
\displaystyle\minimize_{u,z,r,D}
& \displaystyle \frac{\|s\|^2+a^2\|u\|^2+\tr(DD^T)}{\frac{1}{2}\log(1+(s+abu)^Tz)},\\
\text{subject to} 
& Ds=u,\\
& Dr=z-s,\\
& b^2 D^Tz=abs-r,\\
& D_{ij}=0,\quad j\ge i.
\end{array}
\ee

The next step is to write optimality conditions for this optimization problem. Since the problem is non-convex, these are necessary but not sufficient conditions. Specifically, in the following lemma, proved in Appendix~\ref{appendix:optimality_conditions}, we find necessary conditions for optimal $u,z,r,D$ given  $s$. While this lemma is not actually necessary to the proof of Thm.~\ref{thm:main}, we include it as motivation for the system of differential equations that we study in Sec.~\ref{sec:DE}.

\begin{lemma}\label{lemma:optimality_conditions}
    Fix $k\in\bbN$ and any vector $s\in\bbR^k$ such that $\|s\|^2=Q_1>0$. Let $u,z,r,D$ solve \eqref{transformed_optimization}. Let
    \begin{align}
S_i&=\sum_{j:j\le i}s_j^2, &
T_i&=\sum_{j:j\le i}r_js_j,&
R_i&=\sum_{j:j\le i}r_j^2,\label{running_sums1}\\
V_i&=\sum_{j:j> i}u_jz_j, &
Z_i&=\sum_{j:j> i}z_j^2.\label{running_sums2}
\end{align}
    There exists $\lambda> 0$ such that 
     \begin{align}
         D_{ij}&=-a^2 u_is_j+ \frac{z_ir_j}{\lambda},\quad j<i,\\
u_i&=\frac{T_{i-1}s_i}{(1+a^2S_{i-1})(\lambda- R_{i-1})+ a^2 T_{i-1}^2},\label{u_opt}\\
z_i&=\frac{\lambda(1+a^2S_{i-1})s_i}{(1+a^2S_{i-1})(\lambda- R_{i-1})+ a^2 T_{i-1}^2},\label{z_opt}\\
r_i&=\frac{\lambda(ab+a^2b^2V_i)s_i}{\lambda+b^2Z_i},\label{r_opt}\\
&(s+abu)^Tz=Z_0+a^2Q_1-\frac{2aT_{k}}{b}+\frac{R_{k}}{b^2},\\
&a^2\|u\|^2+\tr(DD^T)=\frac{aT_{k}}{b\lambda}-\frac{R_{k}}{b^2\lambda}.
     \end{align}
\end{lemma}

\section{From Optimality Conditions to a Differential Equation}
\label{sec:DE}

Consider a fixed vector $s\in\bbR^k$ and the corresponding running sum $S_i$ defined in \eqref{running_sums1}. From the definitions in \eqref{running_sums1}--\eqref{running_sums2}, the remaining sequences satisfy the difference equations
\begin{align}
    V_i&=V_{i-1}-u_iz_i, & Z_i&=Z_{i-1}-z_i^2,\label{difference_eqn1}\\
    T_i&=T_{i-1}+r_is_i, & R_i&=R_{i-1}+r_i^2.\label{difference_eqn2}
\end{align}
In turn, $u_i$ and $z_i$ depend on $T_{i-1}$ and $R_{i-1}$, and $r_i$ depends on $V_{i},Z_{i}$. Thus, these sequences can be computed in order. Moreover, these computations can be viewed as a variation on Euler's method to approximate the solution to a differential equation. This differential equation emerges from Lemma~\ref{lemma:optimality_conditions} by taking a limit as $\max_i s_i^2\to 0$. Since $S_0=0$ and $S_k=\|s\|^2=Q_1$, we use $S$ as the independent variable, where $S$ goes from $0$ to $Q_1$. We use $(\cdot)'$ for $\frac{d}{dS}(\cdot)$. The system of differential equations is
\begin{align}
u&=\frac{T}{(1+a^2S)(\lambda-R)+a^2T^2},\label{DE1}\\
z&=\frac{\lambda(1+a^2S)}{(1+a^2S)(\lambda-R)+a^2T^2},\\
r&=\frac{\lambda(ab+a^2b^2V)}{\lambda+b^2Z},\\
T'&=r,\quad
R'=r^2,\quad
V'=-uz,\quad
Z'=-z^2.\label{DE7}
\end{align}
We also have the boundary conditions
\be\label{DE_boundary}
T(0)=R(0)=V(Q_1)=Z(Q_1)=0.
\ee

In order to show that this differential equation does in fact result from taking a limit of the optimality conditions in Lemma~\ref{lemma:optimality_conditions}, we need a very general lemma regarding a variation on Euler's method for approximating a differential equation. Specifically, consider an arbitrary multi-dimensional differential equation
\be\label{general_DE}
\theta'(t)=F(\theta(t),t)
\ee
where $\theta(t)\in\bbR^m$. The usual Euler's method is to fix a step size $\Delta>0$, and form a sequence by initializing $\theta_0=\theta(0)$, and then proceeding forward by
\be\label{standard_Euler}
\theta_{i+1}=\theta_i+\Delta F(\theta_i,i\Delta).
\ee
A careful consideration of \eqref{difference_eqn1}--\eqref{difference_eqn2} reveals that this is not quite what is happening there, because the updates of $T_i$ and $R_i$ use $r_i$, which is a function of the just-calculated values of $V_i,Z_i$, rather than those from the previous round, $V_{i-1},Z_{i-1}$. Thus, we need to analayze a  slight variation on Euler's method, wherein the entries of $\theta$ are updated one-by-one, each using the most recent value of the other entries when calculating the next step. Specifically, for any $j\in\{0,\ldots,m\}$, let
\be\label{theta_ij}
\theta_i^{(j)}=(\theta_{i+1,1},\ldots,\theta_{i+1,j-1},\theta_{i,j},\ldots,\theta_{i,m}).
\ee
Now we replace the update rule in \eqref{standard_Euler} by
\be\label{modified_Euler}
\theta_{i+1,j}=\theta_{i,j}+\Delta F_j(\theta_i^{(j)},i\Delta),\quad j=1,\ldots,m.
\ee
The following lemma, proved in Appendix~\ref{appendix:Euler}, shows that, for any differential equation that is sufficiently well-behaved, this modified Euler's method approaches the true solution as the step size goes to zero. The proof of the lemma is similar to the proof that the global error of the standard Euler's method vanishes with step size, as in \cite[Thm.~12.2]{introductionSuli2003}.

\begin{lemma}\label{lemma:Euler}
    Let $\theta(t)\in\bbR^m$ for $t\in[0,T]$ satisfy the differential equation in \eqref{general_DE}, and let $\theta_0,\theta_1,\ldots$ be derived from the update rule \eqref{modified_Euler} with initial condition $\theta_0=\theta(0)$ and step size $\Delta$.
    Assume there exist sets $\calA(t)\subset\bbR^m$ for $t\in[0,T]$ such that
    \be\label{eta_def}
    \inf\{\|\theta(t)-\theta\|:t\in[0,T],\theta\in\bbR^m\setminus\calA(t)\}>0.
    \ee
    Also, assume there exist constants $K_1,K_2,K_3$ such that, for all $t\in[0,T]$, $\theta,\tilde\theta\in \calA(t)$, and all $j\in\{1,\ldots,m\}$
    \begin{align}
    |F_j(\theta,t)-F_j(\tilde\theta,t)|&\le K_1\|\theta-\tilde\theta\|,\label{K1_bound}\\
    \left|\frac{\partial}{\partial t} F_j(\theta,t)\right|&\le K_2,\label{K2_bound}\\
    \|F(\theta,t)\|&\le K_3.\label{K3_bound}
    \end{align}
    Then
    \be
    \lim_{\Delta\to 0}\ \max_{i=0,\ldots,\floor{T/\Delta}} \|\theta(i\Delta)-\theta_i\| =0.
    \ee
\end{lemma}

The following lemma, proved in Appendix~\ref{appendix:DE}, shows that our differential equation of interest in \eqref{DE1}--\eqref{DE_boundary} does in fact satisfy the assumptions of Lemma~\ref{lemma:Euler}, at least under certain conditions, and thus the sequences defined by the optimality conditions in Lemma~\ref{lemma:optimality_conditions} do approach the differential equation solutions, which leads to an achievable energy-per-bit.

\begin{lemma}\label{lemma:DE}
    Fix any $\lambda>0$ and $Q_1>0$. Let $u,z,r,T,R,V,Z$ be functions that solve the differential equation system \eqref{DE1}--\eqref{DE_boundary} where
    \begin{align}
\inf_{S\in[0,Q_1]} (1+a^2S)(\lambda-R(S))+a^2 T(S)^2&>0,
    \label{DE_assumption}
\\
\sup_{S\in [0,Q_1]} \max\{|T(S)|,|R(S)|,|V(S)|,|Z(S)|\}&<\infty.\label{DE_bounding_assumption}
\end{align}
    Then
    \be
\calE^*_{\text{1LR}}\le \frac{Q_1+\frac{aT(Q_1)}{b\lambda}-\frac{R(Q_1)}{b^2\lambda}}{\frac{1}{2}\log\left(1+Z(0)+a^2Q_1-\frac{2aT(Q_1)}{b}+\frac{R(Q_1)}{b^2}\right)}.
    \ee
\end{lemma}

To proceed, we analyze the the differential equation system in \eqref{DE1}--\eqref{DE7}. Define the following variations on $T,R,Z,V$:
\begin{align}
\barT&=\frac{T}{\lambda}, &
\barR&=\frac{R-\lambda}{\lambda^2},\label{bar_defs1}
\\
 \barV&=a^2\lambda^2\left(\frac{1}{ab}+V\right), &
 \barZ&=\lambda^2\left(\frac{\lambda}{b^2}+Z\right). 
 \label{bar_defs2}
\end{align}
Furthermore, let $\barS=1/a^2+S$ (so that derivatives with respect to $S$ are equivalent to derivatives with respect to $\barS$). Then the system simplifies to
\begin{align}
\barT'&=\frac{\barV}{\barZ},
& \barR'&=(\barT')^2,
\\ \barZ'&=-\left(\frac{\barS}{-\barS\barR+\barT^2}\right)^2,
& \barV'&=\frac{\barT\barZ'}{\barS},
\end{align}
with boundary conditions
\begin{align}
\barT(0)&=0,
& \barR(0)&=-\frac{1}{\lambda},\label{bar_boundary1}
\\ \barZ(Q_1)&=\frac{\lambda^3}{b^2},
& \barV(Q_1)&=\frac{a\lambda^2}{b}.\label{bar_boundary2}
\end{align}
Observe that
\begin{align}
(\barS\barV-\barT\barZ)'&=\barV+\barS\barV'-\barT'\barZ-\barT\barZ'
\\&=\barV+\barT\barZ'-\barV-\barT\barZ'
=0.
\end{align}
Thus, for any solution, there exists a constant $c_1$ where\footnote{We write $c_1^3$ instead of $c_1$ because later on it will make things simpler.}
\be\label{c1_def}
\barS\barV-\barT\barZ=c_1^3.
\ee

Now, define some new variables (with a lot of hindsight)
\begin{align}
A&=c_1^{2}\frac{\barT^2-\barS\barR}{\barS^2},&
B&=c_1^{-4}\barS\barZ.\label{ABdef}
\end{align}
We have
\begin{align}
A'&=c_1^{2}\left(\frac{2\barT\barT'-\barR-\barS\barR'}{\barS^2}-\frac{2(\barT^2-\barS\barR)}{\barS^3}\right)
\\&=c_1^{2}\frac{2\barS \barT\barT'-\barS\barR-\barS^2(\barT')^2-2T^2+2\barS\barR}{\barS^3}
\\&=c_1^{2}\frac{-(\barT-\barS \barT')^2-T^2+\barS\barR}{\barS^3}
\\&=-\frac{c_1^{2}}{\barS}\left(\frac{\barT}{\barS}-\frac{\barV}{\barZ}\right)^2-\frac{A}{\barS}
\\&=-\frac{c_1^{2}}{\barS}\left(\frac{\barT\barZ-\barS\barV}{\barS \barZ}\right)^2-\frac{A}{\barS}
\\&=-\frac{c_1^8}{\barS^3\barZ^2}-\frac{A}{\barS}
\\&=-\frac{1}{\barS B^2}-\frac{A}{\barS}.
\end{align}
Also
\begin{align}
B'&=c_1^{-4}(\barS\barZ'+\barZ)
\\&=c_1^{-4}\left(-\barS\left(\frac{\barS}{-\barS\barR+\barT^2}\right)^2+\barZ\right)
\\&=-\frac{c_1^{-4}\barS^3}{(-\barS\barR+\barT^2)^2}+c_1^{-4}\barZ
\\&=-\frac{1}{\barS A^2}+\frac{B}{\barS}.
\end{align}
That is, we have the self-contained two-dimensional system
\be\label{2d_system}
A'=-\frac{1}{\barS B^2}-\frac{A}{\barS},\qquad B'=-\frac{1}{\barS A^2}+\frac{B}{\barS}.
\ee
To solve the two-dimensional system, we observe that
\begin{align}
&\left(AB+\frac{1}{A}-\frac{1}{B}\right)'
=\left(B-\frac{1}{A^2}\right)A'+\left(A+\frac{1}{B^2}\right)B'
\\&=\left(B-\frac{1}{A^2}\right)\left(-\frac{1}{\barS B^2}-\frac{A}{\barS }\right)
\nonumber\\&\qquad
+\left(A+\frac{1}{B^2}\right)\left(-\frac{1}{\barS A^2}+\frac{B}{\barS }\right)=0.
\end{align}
Thus, there exists a constant $c_2$ where
\be\label{AB_eqn}
AB+\frac{1}{A}-\frac{1}{B}=c_2.
\ee
Note that $B=f(A)$ solves \eqref{AB_eqn}, where $f$ is defined in \eqref{f_def} with $\phi=c_2$. Assuming $B=f(A)$, we can reduce to a single-variable differential equation
\be\label{DE_for_A}
A'=\frac{1}{1/a^2+S}\left(-\frac{1}{f(A)^2}-A\right).
\ee
To complete the achievability proof in Thm.~\ref{thm:main}, we fix $A_f,B_f>0$, and construct a solution to the two-dimensional system where  $A(Q_1)=A_f$, $B(Q_1)=B_f$, with $c_1=b\psi$ and $c_2=\phi$. Given these solutions, we can construct solutions to the original system that satisfy the boundary constraints. The following lemma, proved in Appendix~\ref{appendix:final_achievability}, provides the details, showing that these 
differential equation solutions exist, that they satisfy the assumptions of Lemma~\ref{lemma:DE}, and giving the result in the theorem.

\begin{lemma}\label{lemma:final_achievability}
    Given any $A_f,B_f>0$ satisfying $A_f/B_f\le a^2$, let $(A_0,c_1)$ be the pair from Lemma~\ref{lemma:A0c1}. Then there exist $\lambda>0$ and functions that solve the differential equation system \eqref{DE1}--\eqref{DE_boundary}, satisfying the conditions \eqref{DE_assumption}--\eqref{DE_bounding_assumption}, where $Q_1$ is given in \eqref{Q1_thm}, 
    \be
\frac{aT(Q_1)}{b\lambda}-\frac{R(Q_1)}{b^2\lambda}=Q_2
    \ee
    where $Q_2$ is given in \eqref{Q2_thm}, and
    \begin{align}
1+Z(0)+a^2Q_1-\frac{2aT(Q_1)}{b}+\frac{R(Q_1)}{b^2}
\\=\frac{A_0}{a^2}\left(\frac{1}{B_f}+A_0B_0-A_fB_f\right).
    \end{align}
\end{lemma}

\section*{Acknowledgments}

The authors would like to thank Cynthia Keeler for her help solving the 4-dimensional system of differential equations.
This work is supported in part by NSF grants CCF-1817241, CCF-1908725, CCF-1909451, CCF-2107526, and CCF-2245204.

\bibliographystyle{IEEEtran}
\bibliography{relaychannel}

\begin{thebibliography}{1}
\providecommand{\url}[1]{#1}
\csname url@samestyle\endcsname
\providecommand{\newblock}{\relax}
\providecommand{\bibinfo}[2]{#2}
\providecommand{\BIBentrySTDinterwordspacing}{\spaceskip=0pt\relax}
\providecommand{\BIBentryALTinterwordstretchfactor}{4}
\providecommand{\BIBentryALTinterwordspacing}{\spaceskip=\fontdimen2\font plus
\BIBentryALTinterwordstretchfactor\fontdimen3\font minus
  \fontdimen4\font\relax}
\providecommand{\BIBforeignlanguage}[2]{{%
\expandafter\ifx\csname l@#1\endcsname\relax
\typeout{** WARNING: IEEEtran.bst: No hyphenation pattern has been}%
\typeout{** loaded for the language `#1'. Using the pattern for}%
\typeout{** the default language instead.}%
\else
\language=\csname l@#1\endcsname
\fi
#2}}
\providecommand{\BIBdecl}{\relax}
\BIBdecl

\bibitem{Shannon:49}
C.~E. Shannon, ``Communication in the presence of noise,'' \emph{Proceedings of
  the IRE}, vol.~37, no.~1, pp. 10--21, 1949.

\bibitem{PolyanskiyP:11}
P.~Polyanskiy, H.~V.~P. Poor, and S.~Verd\'u, ``Minimum energy to send $k$ bits
  through the {G}aussian channel with and without feedback,'' \emph{IEEE
  Transactions on Information Theory}, vol.~57, no.~8, pp. 4880 -- 4902, 2011.

\bibitem{Host-Madsen:13}
A.~Host-Madsen, ``Minimum energy per bit in broadcast and interference channels
  with correlated information,'' \emph{IEEE Transactions on Information
  Theory}, vol.~59, no.~6, pp. 3796--3810, 2013.

\bibitem{boundsElGamal2006}
A.~El~Gamal, M.~Mohseni, and S.~Zahedi, ``Bounds on capacity and minimum
  energy-per-bit for {AWGN} relay channels,'' \emph{IEEE Transactions on
  Information Theory}, vol.~52, no.~4, pp. 1545--1561, 2006.

\bibitem{CapacityCover1979}
T.~Cover and A.~Gamal, ``Capacity theorems for the relay channel,'' \emph{IEEE
  Transactions on Information Theory}, vol.~25, no.~5, pp. 572--584, 1979.

\bibitem{JointKim2012}
C.~Kim, Y.~Sung, and Y.~H. Lee, ``A joint time-invariant filtering approach to
  the linear {Gaussian} relay problem,'' \emph{IEEE Transactions on Signal
  Processing}, vol.~60, no.~8, pp. 4360--4375, 2012.

\bibitem{JointGohary2013}
R.~H. Gohary and H.~Yanikomeroglu, ``Joint optimization of the transmit
  covariance and relay precoder in general {Gaussian} amplify-and-forward relay
  channels,'' \emph{IEEE Transactions on Information Theory}, vol.~59, no.~9,
  pp. 5331--5351, 2013.

\bibitem{introductionSuli2003}
E.~Süli and D.~F. Mayers, \emph{An Introduction to Numerical Analysis}.\hskip
  1em plus 0.5em minus 0.4em\relax Cambridge University Press, 2003.

\end{thebibliography}

\appendices

\section{Proof of Lemma~\ref{lemma:A0c1}}\label{appendix:A0c1}

From \eqref{second_integral_eqn}, we have
\begin{align}
&\ln\left(\frac{A_0^2}{\psi^2}\right)
\\&=\ln\left(\frac{a^4}{B_f}\right)-\ln A_0+\int_{A_f}^{A_0}\frac{f(w)^2\,dw}{1+wf(w)^2}
\\&=\ln\left(\frac{a^4}{B_f}\right)-\ln A_f-\int_{A_f}^{A_0}\frac{dw}{w}+\int_{A_f}^{A_0}\frac{f(w)^2\,dw}{1+wf(w)^2}
\\&=\ln\left(\frac{a^4}{A_f B_f}\right)-\int_{A_f}^{A_0}\left(\frac{1}{w}-\frac{f(w)^2}{1+wf(w)^2}\right)dw.\label{negative_integrand}
\end{align}
Assuming $A_0,\psi>0$, this can be written
\be
\frac{A_0}{\psi}=\frac{a^2}{\sqrt{A_fB_f}}\exp\left(-\frac{1}{2}\int_{A_f}^{A_0}\left(\frac{1}{w}-\frac{f(w)^2}{1+wf(w)^2}\right)dw\right).
\ee
Plugging this expression for $A_0/\psi$ into \eqref{first_integral_eqn}, we see that \eqref{first_integral_eqn}--\eqref{second_integral_eqn} are solved if and only if $g(A_0)=0$, where
\begin{align}
g(A_0)&=\frac{1}{B_f}+\int_{A_f}^{A_0} \frac{f(w)\,dw}{1+wf(w)^2}\nonumber\\
&-\frac{a}{\sqrt{A_fB_f}}
\exp\left(-\frac{1}{2}\int_{A_f}^{A_0}\left(\frac{1}{w}-\frac{f(w)^2}{1+wf(w)^2}\right)dw\right).\label{zero_function}
\end{align}
We first show that $g(A_0)$ is strictly increasing in $A_0$ for all $A_0>0$. To do so, it is sufficient to show that both integrands in \eqref{zero_function} are strictly positive for all $w>0$. For any $w>0$, $f(w)$ is finite and positive. This implies that the first integrand in \eqref{zero_function} is positive. In addition, 
\be
\frac{1}{w}-\frac{f(w)^2}{1+wf(w)^2}
=\frac{1}{w}-\frac{1}{1/f(w)^2+w}
> \frac{1}{w}+\frac{1}{w}=0.
\ee
This shows that the second integrand in \eqref{zero_function} is positive for all $w>0$.

Since we have shown that $g(A_0)$ is strictly increasing in $A_0$, to see that there exists $A_0\ge A_f$ satisfying $g(A_0)=0$, it is enough to show that $g(A_f)\le 0$ and that $\lim_{A_0\to\infty} g(A_0)>0$. To show the first, note that
\be
g(A_f)=-\frac{a}{\sqrt{A_fB_f}}+\frac{1}{B_f}=\frac{-a+\sqrt{A_f/B_f}}{\sqrt{A_fB_f}}.
\ee
Thus, by the assumption that $A_f/B_f\le a^2$, $g(A_f)\le 0$.

Now consider taking a limit as $A_0\to\infty$. For any $c_2$, as $w\to\infty$,
\be
f(w)=\frac{1}{\sqrt{w}}+O\left(\frac{1}{w}\right),
\ee
so we can write each of the integrands as
\begin{align}
\frac{f(w)^2}{1+w f(w)^2}&=O\left(\frac{1}{w}\right),\\
\frac{f(w)}{1+w f(w)^2}&=O\left(\frac{1}{\sqrt{w}}\right).
\end{align}
Thus, in the limit as $A_0\to \infty$,
\begin{align}
g(A_0)&=\int_{A_f}^{A_0}O\left(\frac{1}{w}\right) dw
\nonumber
\\&\qquad-O(1)\exp\left(-\int_{A_f}^{A_0}O\left(\frac{1}{\sqrt{w}}\right)dw\right)
\\&=O(\log A_0)-O(1)\exp\left(-O\left(\sqrt{A_0}\right)\right).
\end{align}
This shows that $\lim_{A_0\to\infty} g(A_0)=\infty$.

\section{Proof of Lemma~\ref{lemma:optimality_conditions}}\label{appendix:optimality_conditions}

Let
\begin{align}
    g_1&=\|s\|^2+a^2\|u\|^2+\tr(DD^T),\\
    g_2&=(s+abu)^Tz.
\end{align}
For any fixed nonzero $s$, finite minimum energy per bit is achievable, so we may assume that for any optimal $u,z,r,D$, the denominator in the objective function in \eqref{transformed_optimization} is positive, which means $g_2>0$. Similarly, since $s$ is assumed to be nonzero, $g_1>0$. The gradient of the objective function has the form
\begin{align}
&\nabla \frac{g_1}{\frac{1}{2}\log(1+g_2)}
\\&=\frac{\nabla g_1}{\frac{1}{2}\log(1+g_2)}
-\frac{g_1}{\left(\frac{1}{2}\log(1+g_2)\right)^2}\frac{\log e}{2(1+g_2)}\nabla g_2
\\&=\kappa_1\nabla g_1-\kappa_2\nabla g_2
\end{align}
where, from the above arguments, $\kappa_1,\kappa_2>0$.

Even though our optimization problem is not convex, the KKT conditions constitute necessary conditions for any optimal point. The Lagrangian is given by (where we are ignoring the constraints that $D$ is lower triangular, since we will only consider $D_{ij}$ for $j<i$ to be optimization variables)
\begin{align}
L&=\frac{g_1}{\frac{1}{2}\log(1+g_2)}+\nu_1^T(Ds-u)+\nu_2^T(Dr-z+s)\nonumber
\\&\qquad+\nu_3^T(b^2 D^Tz-abs+ r)
\end{align}
where $\nu_1,\nu_2,\nu_3$ are dual variables for the three equality constraints. Optimality conditions are found by differentiating with respect to each variable:
\begin{align}
\frac{\partial L}{\partial D_{ij}}&=2\kappa_1 D_{ij}+\nu_{1i}s_j+\nu_{2i} r_j+b^2z_i\nu_{3j}=0,\label{opt1D}\\
\nabla_u L&=2\kappa_1 a^2 u-\kappa_2 abz-\nu_1=0,\label{opt1u}\\
\nabla_z L&=-\kappa_2(s+abu)-\nu_2+b^2 D\nu_3=0,\label{opt1z}\\
\nabla_r L&= D^T\nu_2+\nu_3=0.\label{opt1r}
\end{align}
Let us solve these equations for $\nu_1,\nu_2,\nu_3$. From \eqref{opt1r}, we have
\be
\nu_3=-D^T\nu_2.
\ee
Plugging this into \eqref{opt1z} gives
\be
-\kappa_2(s+abu)-\nu_2-b^2DD^T\nu_2=0.
\ee
Thus, recalling the definition of $z$,
\be
\nu_2=-\kappa_2(I+b^2DD^T)^{-1}(s+abu)= -\kappa_2z.
\ee
Now
\be
\nu_3=-D^T\nu_2=\kappa_2 D^Tz=\frac{\kappa_2}{b^2}(abs-r).
\ee
From \eqref{opt1u}, we have
\be
\nu_1= 2\kappa_1 a^2 u-\kappa_2 abz.
\ee
From \eqref{opt1D}, we can now write for $j<i$ that
\begin{align}
&2\kappa_1 D_{ij}=-\nu_{1i}s_j-\nu_{2i}r_j-b^2 z_i\nu_{3j}
\\&=(-2\kappa_1 a^2u_i+\kappa_2 abz_i)s_j
+\kappa_2 z_i r_j
+\kappa_2 z_i(-abs_j+r_j)
\\&=-2\kappa_1 a^2 u_is_j+2\kappa_2z_ir_j.
\end{align}
Let $\lambda=\frac{\kappa_1}{\kappa_2}$. Then certainly $\lambda>0$, and
\be
D_{ij}=\begin{cases}-a^2 u_is_j+\frac{z_ir_j}{\lambda}, & j<i\\ 0, & j\ge i.\end{cases}
\ee
The equation $Ds=u$ can be written
\begin{align}
u_i&=\sum_{j:j<i} D_{ij}s_j
\\&=\sum_{j:j<i} \left(-a^2 u_i s_j^2+\frac{1}{\lambda}z_i r_js_j\right)
\\&=-a^2 u_i S_{i-1}+\frac{1}{\lambda}z_i T_{i-1}.\label{Ds_rewrite}
\end{align}
The equation $Dr=z-s$ can be written
\begin{align}
z_i-s_i&=\sum_{j:j<i} D_{ij}r_j
\\&=\sum_{j:j<i}\left(-a^2 u_is_jr_j+\frac{1}{\lambda}z_i r_j^2\right)
\\&=-a^2 u_i T_{i-1}+ \frac{1}{\lambda}z_i R_{i-1}.\label{Dr_rewrite}
\end{align}
The equation $D^Tz=\frac{abs- r}{b^2}$ can be written
\begin{align}
\frac{abs_j- r_j}{b^2}&=\sum_{i:i>j} D_{ij}z_i
\\&=\sum_{i:i>j} \left(-a^2 s_j u_iz_i+\frac{1}{\lambda} r_jz_i^2\right)
\\&=-a^2 s_jV_j+\frac{1}{\lambda}r_jZ_j.\label{Dz_rewrite}
\end{align}
Solving \eqref{Ds_rewrite}, \eqref{Dr_rewrite}, and \eqref{Dz_rewrite} for $u_i,z_i,r_i$ gives \eqref{u_opt}--\eqref{r_opt}.

Recalling that $s+abu=(I+b^2DD^T)z$, we have
\begin{align}
    (s+abu)^Tz&=z^T(I+b^2DD^T)z
    \\&=\|z\|^2+b^2\|Dz\|^2
    \\&=\|z\|^2+\frac{1}{b^2}\|abs-r\|^2
    \\&=\sum_{i=1}^k\left(z_i^2+\frac{1}{b^2}(a^2b^2s_i^2-2abs_ir_i+r_i^2)\right)
    \\&=Z_0+a^2Q_1-\frac{2aT_{k}}{b}+\frac{R_{k}}{b^2}.
\end{align}
We also have
\begin{align}
     &a^2\|u\|^2+\tr(DD^T)
     \\&=a^2\|u\|^2+\sum_{i,j:j<i} D_{ij}^2
     \\&=a^2\|u\|^2+\sum_{i,j:j<i} D_{ij}\left(-a^2 u_is_j+\frac{z_ir_j}{\lambda}\right)
     \\&=a^2\|u\|^2-a^2\sum_i u_i \sum_{j:j<i}D_{ij}s_j+\frac{1}{\lambda} \sum_j r_j \sum_{i:j<i}D_{ij}z_i
     \\&=a^2\|u\|^2-a^2\sum_i u_i^2+\frac{1}{\lambda} \sum_{j} r_j\frac{abs_j-r_j}{b^2}
     \\&=\frac{aT_{k}}{b\lambda}-\frac{R_{k}}{b^2\lambda}.
\end{align}

\section{Proof of Lemma~\ref{lemma:Euler}}\label{appendix:Euler}

Let $\eta$ be the LHS of \eqref{eta_def}, so $\eta>0$. Moreover,
\be\label{eta_to_calA}
\theta\in\calA(t),\text{ if }\|\theta(t)-\theta\|<\eta.
\ee

For any $t$ and $j\in\{0,\ldots,m\}$, let
\be
\theta^{(j)}(t)=(\theta_1(t+\Delta),\ldots,\theta_{j-1}(t+\Delta),\theta_j(t),\ldots,\theta_m(t)).
\ee
Note that $\theta^{(m)}(t)=\theta^{(0)}(t+\Delta)$; similarly, by the definition in \eqref{theta_ij}, $\theta_i^{(m)}=\theta_{i+1}^{(0)}$. Now define $\eps_i^{(j)}=\theta^{(j)}(i\Delta)-\theta_i^{(j)}$. As above, $\eps_i^{(m)}=\eps_{i+1}^{(0)}$. Since we initialize the Euler approximation with $\theta_0=\theta(0)$, and $\theta_0^{(0)}=\theta_0$, we have $\eps_0^{(0)}=0$. For any $i$ and any $j\in\{0,\ldots,m-1\}$, we compare $\eps_i^{(j)}$ to $\eps_i^{(j+1)}$. Note that these vectors differ only in their $j$th entries, so
\begin{align}
&\|\eps_{i}^{(j+1)}-\eps_i^{(j)}\|
\\&=|\theta_j((i+1)\Delta)-\theta_{i+1,j}-\theta_j(i\Delta)+\theta_{i,j}|
\\&=|\theta_j(i\Delta)-\theta_j((i+1)\Delta)-\Delta F_j(\theta_i^{(j)},i\Delta)|
\\&=\left|\int_{i\Delta}^{(i+1)\Delta} (F_j(\theta(t),t)-F_j(\theta_i^{(j)},i\Delta))dt\right|
\\&\le \int_{i\Delta}^{(i+1)\Delta} \left|F_j(\theta(t),t)-F_j(\theta_i^{(j)},i\Delta)\right|dt.
\label{Fj_integral}
\end{align}
For any $t\in[i\Delta,(i+1)\Delta]$,
\begin{align}
    &|F_j(\theta(t),t)-F_j(\theta_i^{(j)},i\Delta)|
    \\&\le |F_j(\theta(t),t)-F_j(\theta(t),i\Delta)|+|F_j(\theta(t),i\Delta)-F_j(\theta_i^{(j)},i\Delta)|
    \\&\le \Delta K_2+|F_j(\theta(t),i\Delta)-F_j(\theta_i^{(j)},i\Delta)|.\label{Fj_diff}
\end{align}
The next step is to use the bound in \eqref{K1_bound} on the difference between $F_j$ evaluated at different $\theta$ values. Suppose $\theta_i^{(j)}\in\calA(i\Delta)$. In order to apply \eqref{K1_bound} to \eqref{Fj_diff}, we need $\theta(t)\in\calA(i\Delta)$ as well; but we only know that $\theta(t)\in\calA(t)$. However, by \eqref{K3_bound}, we have
\be
\|\theta(t)-\theta(i\Delta)\|
\le \Delta K_3.
\ee
Thus, if $\Delta$ is small enough so that $\Delta K_3<\eta$, then by \eqref{eta_to_calA}, $\theta(t)\in\calA(i\Delta)$. Now we can apply \eqref{K1_bound} to find
\be\label{K1_application}
|F_j(\theta(t),i\Delta)-F_j(\theta_i^{(j)},i\Delta)|\le K_1\|\theta(t)-\theta_i^{(j)}\|.
\ee
Thus, 
\begin{align}
    &|F_j(\theta(t),t)-F_j(\theta_i^{(j)},i\Delta)|
    \\&\le \Delta K_2+|F_j(\theta(t),i\Delta)-F_j(\theta_i^{(j)},i\Delta)|
    \\&\le \Delta K_2+K_1\|\theta(t)-\theta_i^{(j)}\|
    \\&\le\Delta K_2+K_1\|\theta(t)-\theta^{(j)}(i\Delta)\|+\|\theta^{(j)}(i\Delta)-\theta_i^{(j)}\|
    \\&= \Delta K_2+K_1(\|\theta(t)-\theta^{(j)}(i\Delta)\|+\|\eps_i^{(j)}\|).
\end{align}
To bound the second term, note that
\begin{align}
    &\|\theta(t)-\theta^{(j)}(i\Delta)\|
    \\&\le \|\theta(t)-\theta(i\Delta)\|+\|\theta(i\Delta)-\theta^{(j)}(i\Delta)\| 
    \\&\le \|\theta(t)-\theta(i\Delta)\|+\|\theta(i\Delta)-\theta((i+1)\Delta)\| \label{theta_t_intermediate}
    \\&\le 2\Delta K_3
\end{align}
where in \eqref{theta_t_intermediate} we have used the fact that $\theta^{(j)}(i\Delta)$ equals $\theta((i+1)\Delta)$ in the first $j-1$ entries and $\theta(i\Delta)$ in the remaining entries.
Thus
\be
|F_j(\theta(t),t)-F_j(\theta_i^{(j)},i\Delta)|\le \Delta(K_2+2K_1K_3)+K_1\|\eps_i^{(j)}\|.
\ee
Plugging this bound into the integral in \eqref{Fj_integral} and defining $K=K_2+2K_1K_3$, we have
\be
\|\eps_i^{(j+1)}-\eps_i^{(j)}\|\le \Delta^2K+\Delta K_1\|\eps_i^{(j)}\|
\ee
which means
\begin{align}
\|\eps_i^{(j+1)}\|&\le (1+\Delta K_1)\|\eps_i^{(j)}\|+\Delta^2K
\\&\le e^{\Delta K_1}\|\eps_i^{(j)}\|+\Delta^2K.\label{eps_subsequent_bound}
\end{align}

Assume that $\Delta$ is small enough so that
\be\label{Delta_small_enough}
\frac{\Delta K}{K_1}(e^{K_1mT}-1)+\Delta K_3<\eta.
\ee
Note that this assumption on $\Delta$ also satisfies $\Delta K_3<\eta$, which was used to show \eqref{K1_application}.
We prove by induction that, for all $i\in\{0,\ldots,\floor{T/\Delta}-1\}$ and $j\in\{0,\ldots,m\}$, $\theta_i^{(j)}\in\calA(i\Delta)$ and
\be\label{bound_by_induction}
\|\eps_i^{(j)}\|\le \frac{\Delta K}{K_1}(e^{\Delta K_1(im+j)}-1).
\ee
The base case corresponds to $i=0,j=0$. Indeed $\eps_0^{(0)}=0$, which satisfies \eqref{bound_by_induction}, and also $\theta_0^{(0)}=\theta_0=\theta(0)\in\calA(0)$. Now suppose the induction hypothesis is true for $i$ and some $j<m$. Then, by assumption $\theta_i^{(j)}\in\calA(i\Delta)$, from \eqref{eps_subsequent_bound} we have
\begin{align}
\|\eps_i^{(j+1)}\|&\le e^{\Delta K_1}\frac{\Delta K}{K_1}(e^{\Delta K_1(im+j)}-1)+\Delta^2K
\\&=\frac{\Delta K}{K_1}\left(e^{\Delta K_1(im+j+1)}-e^{\Delta K_1}+\Delta K_1\right)
\\&\le \frac{\Delta K}{K_1}\left(e^{\Delta K_1(im+j+1)}-1\right).
\end{align}
This proves the necessary bound on $\|\eps_i^{(j+1)}\|$. To show that $\theta_i^{(j+1)}\in\calA(i\Delta)$, note that
\begin{align}
\|\theta_i^{(j+1)}-\theta(i\Delta)\|
&\le \|\eps_i^{(j+1)}\|+\|\theta^{(j+1)}(i\Delta)-\theta(i\Delta)\|
\\&\le \frac{\Delta K}{K_1}\left(e^{\Delta K_1(im+j+1)}-1\right)
+\Delta K_3
\\&\le \frac{\Delta K}{K_1}\left(e^{ K_1mT}-1\right)
+\Delta K_3
\\&<\eta
\end{align}
where the last step follows from the assumption on $\Delta$ in  \eqref{Delta_small_enough}. By the fact in \eqref{eta_to_calA} that any $\theta$ close enough to $\theta(t)$ must be in $\calA(t)$, and the assumption that $\theta(i\Delta)\in\calA(i\Delta)$, this shows that $\theta_i^{(j+1)}\in\calA(i\Delta)$. Now suppose the induction hypothesis holds for some $i$ and $j=m$, and we prove it for $i+1$ and $j=0$. Recall that $\eps_i^{(m)}=\eps_{i+1}^{(0)}$, so the bound on $\eps_{i+1}^{(0)}$ in \eqref{bound_by_induction} is immediate. To show that $\theta_{i+1}^{(0)}\in\calA((i+1)\Delta)$, note that $\theta^{(0)}((i+1)\Delta)=\theta((i+1)\Delta)$, so
\begin{align}
    \|\theta_{i+1}^{(0)}-\theta((i+1)\Delta)\|
    &=\|\eps_{i+1}^{(0)}\|
    <\eta
\end{align}
where the last step follows from a similar bound as above. Again using \eqref{eta_to_calA}, this shows that $\theta_{i+1}^{(0)}\in\calA((i+1)\Delta)$, which completes the induction hypothesis.

From \eqref{bound_by_induction}, for 
any $i\in\{1,\ldots,\floor{T/\Delta}\}$
\begin{align}
\|\theta_i-\theta(i\Delta)\|
&=\|\theta^{(m)}_{i-1}-\theta^{(m)}((i-1)\Delta)\|
\\&=\|\eps_{i-1}^{(m)}\|
\\&\le \frac{\Delta K}{K_1}(e^{\Delta K_1 im}-1)
\\&\le \frac{\Delta K}{K_1}(e^{ K_1 mT}-1).
\end{align}
Taking a limit as $\Delta\to 0$ proves the lemma.

\section{Proof of Lemma~\ref{lemma:DE}}\label{appendix:DE}

Fix $\lambda,Q_1$, and a solution to the differential equation as stated in the Lemma. Let $\theta(S)=(V(S),Z(S),T(S),R(S))$ be the 4-dimensional solution to the DE. The full system can be written $\theta'=F(\theta,S)$ where
\begin{align}
    F_1(\theta,S)&=-\frac{\lambda \theta_3(1+a^2S)}{((1+a^2S)(\lambda-\theta_4)+a^2\theta_3^2)^2},\\
    F_2(\theta,S)&=-\frac{\lambda^2 (1+a^2S)^2}{((1+a^2S)(\lambda-\theta_4)+a^2\theta_3)^2},\\
    F_3(\theta,S)&=\frac{\lambda(ab+a^2b^2\theta_1)}{\lambda+b^2\theta_2},\\
    F_4(\theta,S)&=F_3(\theta,S)^2.
\end{align}
We wish to apply Lemma~\ref{lemma:Euler}, so we need to verify the assumptions for this particular differential equation and the trajectory satisfying the assumptions in \eqref{DE_assumption}--\eqref{DE_bounding_assumption}. We know there exist constants $\gamma>0$ and $K_0$ where, for all $S\in[0,Q_1]$,
\begin{align}
(1+a^2S)(\lambda-R(S))+a^2 T(S)^2&\ge 2\gamma,\\
\max\{|V(S)|,|Z(S)|,|T(S)|,|R(S)|\}&\le K_0-\gamma.
\end{align}
We define $\calA(S)$ as the set of $\theta\in\bbR^4$ where
\begin{align}
(1+a^2S)(\lambda-\theta_4)+a^2 \theta_3^2&\ge \gamma,\label{gamma_def}\\
\|\theta\|_\infty&\le K_0.\label{A_def2}
\end{align}
We need to show there exist constants $K_1,K_2,K_3$ satisfying \eqref{K1_bound}--\eqref{K3_bound}. Since by \eqref{A_def2}, the variables are all bounded, the only possible problem with these constants existing is if the denominator in one of the $F_j$ expressions goes to 0. Recall the boundary condition $Z(Q_1)=0$; since $Z'\le 0$, we know any solution satisfies $Z(S)\ge 0$ for $S\in[0,Q_1]$. Thus, the denominator in $F_3$ and $F_4$ is at least $\lambda$. Similarly, by \eqref{gamma_def}, the denominators in $F_1$ and $F_2$ are at least $\gamma$. This ensures that the derivative with respect to $S$ is bounded (i.e., $K_2$ exists in \eqref{K2_bound}) and the overall function is bounded (i.e., $K_3$ exists in \eqref{K3_bound}). 

However, ensuring that $F_j$ is Lipshitz in $\theta$ (i.e., $K_1$ exists in \eqref{K1_bound}) requires some additional care. Specifically, we know by \eqref{gamma_def} that the norm of the gradient of $F_1(\theta,S)$ and $F_2(\theta,S)$ are bounded for $\theta\in\calA(S)$. However, this does not quite prove \eqref{K1_bound}, because $\calA(S)$ is not a convex set, since \eqref{gamma_def} is not a convex condition. We need to show there exists a path between any $\theta,\tilde\theta\in\calA(S)$ whose length does not exceed a constant multiple of $\|\theta-\tilde\theta\|$. Since \eqref{gamma_def} depends only on $\theta_3,\theta_4$, we only need to worry about these two values. Consider any $\theta,\tilde\theta\in\calA(S)$. If $0\le \theta_3\le \tilde\theta_3$, then we can form the following path from $(\theta_3,\theta_4)$ to $(\tilde\theta_3,\tilde\theta_4)$ composed of two orthogonal straight lines:
\be
(\theta_3,\theta_4)\to (\tilde\theta_3,\theta_4)\to (\tilde\theta_3,\tilde\theta_4).
\ee
This path stays entirely within $\calA(S)$, and its length is
\be
|\theta_3-\tilde\theta_3|+|\theta_4-\tilde\theta_4|\le \|\theta-\tilde\theta\|_1\le 2\|\theta-\tilde\theta\|.
\ee
Essentially the same argument applies whenever $\theta_3,\tilde\theta_3$ have the same sign. Now suppose $\theta_3\le 0\le \tilde\theta_3$. By \eqref{A_def2}, $\theta_3\ge -K_0$. Let $\theta_4^*=\lambda-\frac{\gamma}{1+a^2S}$; thus for any $\theta_4\le \theta_4^*$ and any $\theta_3$, \eqref{gamma_def} automatically holds. We now form the following path:
\begin{multline}
(\theta_3,\theta_4)\to (\theta_3,\min\{\theta_4,\tilde\theta_4,\theta_4^*\})\\
\to (\tilde\theta_3,\min\{\theta_4,\tilde\theta_4,\theta_4^*\})
\to (\tilde\theta_3,\tilde\theta_4).
\end{multline}
Once again this path stays within $\calA(S)$. If $\theta_4\le \theta_4^*$ or $\tilde\theta_4\le \theta_4^*$, then the length of this path is $|\theta_3-\tilde\theta_3|+|\theta_4-\tilde\theta_4|$. On the other hand, if $\theta_4,\tilde\theta_4>\theta_4^*$, then the length is
\be
|\theta_4-\theta_4^*|+|\theta_3-\tilde\theta_3|+|\theta_4^*-\tilde\theta_4|.
\ee
Be rearranging \eqref{gamma_def}, we know
\be
\theta_4\le \lambda+\frac{a^2\theta_3^2-\gamma}{1+a^2S}.
\ee
Thus
\begin{align}
    \theta_4-\theta_4^*
    &\le \frac{a^2\theta_3^2}{1+a^2S}
    \\&\le a^2\theta_3^2
    \\&\le a^2K_0 (-\theta_3)
    \\&\le a^2K_0(\tilde\theta_3-\theta_3)
    \\&=a^2K_0|\tilde\theta_3-\theta_3|.
\end{align}
Now we may bound the length of the path by
\begin{align}
    &|\theta_4-\theta_4^*|+|\theta_3-\tilde\theta_3|+|\theta_4^*-\tilde\theta_4|
    \\&=|\theta_3-\tilde\theta_3|
    +\theta_4+\tilde\theta_4-2\theta_4^*
    \\&=|\theta_3-\tilde\theta_3|+|\tilde\theta_4-\theta_4|+2(\theta_4-\theta_4^*)
    \\&\le(1+2a^2K_0)|\theta_3-\tilde\theta_3|+|\tilde\theta_4-\theta_4|
    \\&\le (1+2a^2K_0)\|\theta-\tilde\theta\|_1
    \\&\le 2(1+2a^2K_0)\|\theta-\tilde\theta\|.
\end{align}
This shows that for any $\theta,\tilde\theta\in\calA(S)$, there exists a path between them whose length does not exceed a constant multiple of $\|\theta-\tilde\theta\|$, which in turn proves \eqref{K1_bound}.

We have now shown that the assumptions of Lemma~\ref{lemma:Euler} hold for our differential equation. Fix $k$, and let $\Delta=\frac{Q_1}{k}$. Let $\theta_0,\ldots,\theta_k$ be the sequence created by the modified Euler method for this differential equation, initialized by $\theta_0=(V(0),Z(0),T(0),R(0))$. By Lemma~\ref{lemma:Euler}, we can conclude that, as $k\to\infty$, $\Delta\to 0$, so
\be
\theta_k\to \theta(Q_1).
\ee
We now form a linear code based on the $\theta_i$ sequence as follows. Let $S_i=i\Delta$ for $i=0,\ldots,k$ and $s_i=\sqrt{\Delta}$ for $i=1,\ldots,k$. Also let
\begin{align}
V_i&=\theta_{i,1},&Z_i&=\theta_{i,2},&T_i&=\theta_{i,3},&R_i&=\theta_{i,4}.
\end{align}
Furthermore, define $u_i,z_i,r_i$ for $i=1,\ldots,k$ according to the optimality conditions from the finite-dimensional optimization problem in \eqref{u_opt}--\eqref{r_opt}. 

The above construction nearly satisfies all the optimality conditions from Lemma~\ref{lemma:optimality_conditions}; however, as we will show, $V_i,Z_i$ are not quite the right-sided running sums as in \eqref{running_sums2}.
According to the update rule \eqref{modified_Euler}, we have
\begin{align}
    V_{i+1}&=V_i+\Delta F_1(V_i,Z_i,T_i,R_i,i\Delta)
    \\&=V_i-s_i^2 \frac{\lambda T_i(1+a^2S_i)}{((1+a^2S_i)(\lambda-R_i)+a^2T_i^2)^2}
    \\&=V_i-u_{i+1}z_{i+1}.
\end{align}
Next, we have
\begin{align}
    Z_{i+1}&=Z_i+\Delta F_2(V_{i+1},Z_i,T_i,R_i,i\Delta)
    \\&=Z_i-s_i^2 \frac{\lambda^2 (1+a^2S_i)^2}{((1+a^2S_i)(\lambda-R_i)+a^2T_i)^2}
    \\&=Z_i-z_{i+1}^2.
\end{align}
Next,
\begin{align}
    T_{i+1}&=T_i+\Delta F_3(V_{i+1},Z_{i+1},T_i,R_i,i\Delta)
    \\&=T_i+s_{i+1}^2 \frac{\lambda(ab+a^2b^2V_{i+1})}{\lambda+b^2Z_{i+1}}
    \\&=T_i+r_{i+1}s_{i+1}.
\end{align}
Finally,
\begin{align}
    R_{i+1}&=R_i+\Delta F_3(V_{i+1},Z_{i+1},T_{i+1},R_i,i\Delta)
    \\&=R_i+s_{i+1}^2 \left(\frac{\lambda(ab+a^2b^2V_{i+1})}{\lambda+b^2Z_{i+1}}\right)^2
    \\&=R_i+r_{i+1}^2.
\end{align}
Recalling that $T_0=T(0)=0$ and $R_0=R(0)=0$, we may summarize the above as
\begin{align}
    V_i&=V_0-\sum_{j:j\le i}u_jz_j=V_k+\sum_{j:j>i} u_jz_j,\\
    Z_i&=Z_0-\sum_{j:j\le i} z_j^2=Z_k+\sum_{j:j>i} z_j^2,\\
    T_i&=\sum_{j:j\le i} r_js_j,\\
    R_i&=\sum_{j:j\le i} r_j^2.
\end{align}
Note that, for any finite $k$, there is no guarantee that $V_k=0$ and $Z_k=0$, which means the expressions for $V_i,Z_i$ in \eqref{running_sums2} do not hold; although the expressions for $T_,R_i$ in \eqref{running_sums1}--\eqref{running_sums2} do. However, in the limit as $k\to\infty$, $V_k\to 0$ and $Z_k\to 0$, which means the optimality conditions almost hold.

Now we define the $k\times k$ matrix $D$ by
\be
D_{ij}=\begin{cases}-a^2 u_is_j+\frac{z_ir_j}{\lambda}, & j<i\\ 0, & j\ge i.\end{cases}
\ee
We have
\begin{align}
    (Ds)_i&=\sum_{j:j<i}D_{ij}s_j
    \\&=\sum_{j:j<i}\left(-a^2 u_is_j+\frac{z_ir_j}{\lambda}\right)s_j
    \\&=-a^2 u_i S_{i-1}+\frac{z_i T_{i-1}}{\lambda}
    \\&=u_i.
\end{align}
Similarly,
\begin{align}
    (Dr)_i&=\sum_{j:j<i}D_{ij}r_j
    \\&=\sum_{j:j<i}\left(-a^2 u_is_j+\frac{z_ir_j}{\lambda}\right)r_j
    \\&=-a^2u_iT_{i-1}+\frac{z_iR_{i-1}}{\lambda}
    \\&=z_i-s_i.
\end{align}
Thirdly,
\begin{align}
    (D^Tz)_j&=\sum_{i:i>j}D_{ij}z_i
    \\&=\sum_{i:i>j}\left(-a^2 u_is_j+\frac{z_ir_j}{\lambda}\right)z_i
    \\&=-a^2s_j (V_j-V_k)+\frac{r_j}{\lambda}(Z_j-Z_k)
    \\&=\frac{abs_j-r_j}{b^2}+a^2 s_jV_k-\frac{r_j}{\lambda}Z_k.
\end{align}

We can see that two out of three equality conditions in the optimization problem \eqref{transformed_optimization} hold; again the third holds only in the limit as $k\to\infty$. Even so, we can take the $s,D$ that we have constructed and evaluate $\calE(s,D)$, given in \eqref{EsD_formula}. We have
\begin{align}
    &(I+b^2DD^T)z=z+D\left(abs-r+a^2b^2V_ks-\frac{a^2Z_k}{\lambda}r\right)
    \\&=z+abu-(z-s)+a^2b^2V_ku-\frac{a^2Z_k}{\lambda}(z-s)
    \\&=s+abu+a^2b^2V_ku-\frac{a^2Z_k}{\lambda}(z-s).\label{z_inexact}
\end{align}
As we proved in order to apply Lemma~\ref{lemma:Euler}, the values of the derivative functions in the differential equation are bounded along our trajectory. Since the $u_i,z_i,r_i$ values are essentially values of those functions multiplied by $s_i=\sqrt{\Delta}$, there exists a constant $\Gamma$ where 
\be
|u_i|,|z_i|,|r_i|\le \frac{\Gamma}{\sqrt{Q_1}} \sqrt{ \Delta},\quad i=1,\ldots,k.
\ee
Thus,
\be
\|u\|^2,\|z\|^2,\|r\|^2\le k \frac{\Gamma^2}{Q_1} \Delta=\Gamma^2.
\ee
That is, the norm of each of these vectors is at most $\Gamma$. Let 
\be
w=a^2b^2V_ku-\frac{a^2Z_k}{\lambda}(z-s), 
\ee
so
\be
\|w\|\le \left(a^2b^2V_k+\frac{2a^2Z_k}{\lambda}\right)\Gamma.
\ee
In particular, the norm of $w$ goes to $0$ as $k\to\infty$. From \eqref{z_inexact}, we have
\be
s+abu=(I+b^2DD^T)z-w.
\ee
Thus, the value in the denominator of \eqref{EsD_formula} is
\begin{align}
    &s^T(I+abD^T)(I+b^2DD^T)^{-1}(I+abD)s
    \\&=(s+abu)^T(I+b^2DD^T)^{-1}(s+abu)
    \\&=z^T(I+b^2DD^T)z-2z^Tw+w^T(I+b^2DD^T)^{-1}w
    \\&\le \|z\|^2+b^2\|D^Tz\|^2+2\Gamma\|w\|+\|w\|^2\label{PSD_use}
    \\&=\|z\|^2+b^2\left\|\frac{abs-r}{b^2}+a^2V_ks-\frac{Z_k}{\lambda} r\right\|^2+2\Gamma\|w\|+\|w\|^2
    \\&\le \|z\|^2+\frac{1}{b^2}\|abs-r\|^2+\left(a^2V_k+\frac{Z_k}{\lambda}\right)\Gamma\nonumber
    \\&\qquad+2\Gamma\|w\|+\|w\|^2
    \\&=Z_0-Z_k+a^2Q_1-\frac{2aT_k}{b}+\frac{R_k}{b^2}+\left(a^2V_k+\frac{Z_k}{\lambda}\right)\Gamma\nonumber
    \\&\qquad+2\Gamma\|w\|+\|w\|^2
\end{align}
where in \eqref{PSD_use} we have used the fact that $(I+b^2DD^T)^{-1}\preceq I$. For the value in the numerator of \eqref{EsD_formula}, we have $\|s\|^2=Q_1$, and
\begin{align}
    &a^2\|u\|^2+\tr(DD^T)
    \\&=a^2\|u\|^2+\sum_{i,j:j<i} D_{ij}\left(-a^2u_is_j+\frac{z_ir_j}{\lambda}\right)
    \\&=a^2\|u\|^2-a^2\sum_{i}u_i^2\nonumber
    \\&\qquad+\frac{1}{\lambda}\sum_{j} r_j\left(\frac{abs_j-r_j}{b^2}+a^2 s_jV_k-\frac{r_j}{\lambda}Z_k
    \right)
    \\&=\left(\frac{a}{b\lambda}+\frac{a^2V_k}{\lambda}\right)T_k-
    \left(\frac{1}{b^2\lambda}+\frac{Z_k}{\lambda^2}\right)R_k.
\end{align}
Plugging these values into the definition of $\calE_{\text{1LR}}(s,D)$ in \eqref{EsD_formula}, and then taking a limit as $k\to\infty$ completes the lemma, since in the limit
\begin{align}
(V_k,Z_k,T_k,R_k)&\to(V(Q_1),Z(Q_1),T(Q_1),R(Q_1))
\\&=(0,0,T(Q_1),R(Q_1)).
\end{align}

\section{Proof of Lemma~\ref{lemma:final_achievability}}\label{appendix:final_achievability}

We fix $A_f,B_f>0$ where $A_f/B_f\le a^2$, and we seek a solution where $A(Q_1)=A_f$, $B(Q_1)=B_f$. Also define for convenience $A_0=A(0)$, $B_0=B(0)$. 
We may rewrite \eqref{DE_for_A} as
\be
\frac{dA}{dS}=\frac{1}{1/a^2+S}\left(-\frac{1}{f(A)^2}-A\right).
\ee
Rearranging gives
\be\label{simple_DE}
\frac{f(A)^2\,dA}{1+Af(A)^2}=\frac{-dS}{1/a^2+S}.
\ee
Integrating \eqref{simple_DE} from $S=0$ to $S=Q_1$ gives
\be\label{Q1_integral_equation}
\int_{A_f}^{A_0}\frac{f(w)^2\,dw}{1+wf(w)^2}=\int_0^{Q_1}\frac{dS}{1/a^2+S}=\ln(1+a^2Q_1).
\ee
Assuming \eqref{Q1_integral_equation}  is satisfied, since the integrand on the LHS is non-negative and finite for $w\in[A_f,A_0]$, for each $S\in[0,Q_1]$ there exists $A(S)$ where
\be
\int_{A_f}^{A(S)}
\frac{f(w)^2\,dw}{1+wf(w)^2}
=\int_S^{Q_1}\frac{dv}{1/a^2+v}=\ln\left(\frac{1+a^2Q_1}{1+a^2S}\right).
\ee
This solution $A(S)$ satisfies \eqref{simple_DE}. Thus, if we define $B(S)=f(A(S))$ with $\phi=c_2$, then $(A(S),B(S))$ together solve the two-dimensional system \eqref{2d_system}. In particular, $B_0=f(A_0)$.

Now we need to recover from $A,B$ a solution to the original 4-dimensional system. Let
\be
U=\frac{\barT}{\barS}.
\ee
Then
\be
U'=\frac{\barT'}{\barS}-\frac{\barT}{\barS^2}=\frac{\barV}{\barS \barZ}-\frac{\barT}{\barS ^2}
=\frac{\barS \barV-\barT\barZ}{\barS ^2\barZ}=\frac{c_1^3}{\barS ^2\barZ}=\frac{1}{c_1\barS B}.
\ee
We have
\be
\frac{dU}{dA}=\frac{1/(c_1\barS B)}{-1/(\barS B^2)-A/\barS }
=-\frac{ f(A)}{c_1(1+Af(A)^2)}.
\ee
By the boundary condition for $\barT$, $U(0)=\barT(0)/(1/a^2)=0$. Thus
\be
U(S)=\frac{1}{c_1} \int_{A(S)}^{A_0} \frac{f(w)\,dw}{1+wf(w)^2}
\ee
so
\be
\barT(S)=\frac{1/a^2+S}{c_1}\int_{A(S)}^{A_0} \frac{f(w)\,dw}{1+wf(w)^2}.\label{barT_solution}
\ee
Now we can solve for the remaining variables fairly easily. From the definition of $A$ in \eqref{ABdef},
\be\label{barR_solution}
\barR=\frac{\barT^2}{1/a^2+S}-\frac{(1/a^2+S)A}{c_1^2}.
\ee
From the definition of $B$ in \eqref{ABdef}, 
\be\label{barZ_solution}
\barZ=\frac{c_1^4B}{1/a^2+S}.
\ee
From the constant value in \eqref{c1_def}, 
\be\label{barV_solution}
\barV=\frac{c_1^3+\barT\barZ}{1/a^2+S}.
\ee
We can also recover $R,Z,V,T$ by reversing the linear relationships in \eqref{bar_defs1}--\eqref{bar_defs2}.

Now we show how the boundary conditions can be satisfied. The boundary condition for $\barT(0)$ is already satisfied by \eqref{barT_solution}. By \eqref{barR_solution}, the boundary condition on $\barR(0)$ gives
\be
-\frac{1}{\lambda}=\barR(0)=\frac{\barT(0)^2}{1/a^2}-\frac{A_0}{a^2 c_1^2}
=-\frac{A_0}{a^2 c_1^2}.
\ee
This condition is satisfied by solving for $\lambda$:
\be\label{lambda_formula}
\lambda=\frac{a^2c_1^2}{A_0}.
\ee
By \eqref{barZ_solution}, the boundary condition on $\barZ(Q_1)$ gives
\be
\frac{\lambda^3}{b^2}=\barZ(Q_1)=\frac{c_1^4 B_f}{1/a^2+Q_1}.
\ee
This condition is satisfied by solving for $Q_1$, and using the formula for $\lambda$ from \eqref{lambda_formula}:
\be\label{Q1_formula}
Q_1=-\frac{1}{a^2}+\frac{b^2 c_1^4 B_f}{\lambda^3}
=-\frac{1}{a^2}+\frac{b^2 A_0^3 B_f}{a^6 c_1^2}.
\ee
Plugging this formula for $Q_1$ back into \eqref{Q1_integral_equation} gives the requirement
\be\label{c1_integral_eqn1}
\int_{A_f}^{A_0}\frac{f(w)^2\,dw}{1+wf(w)^2}=\ln\left(\frac{b^2 A_0^3 B_f}{a^6 c_1^2}\right).
\ee
Finally, by \eqref{barV_solution}, the boundary condition for $\barV(Q_1)$ gives
\be
\frac{a\lambda^2}{b}=\barV(Q_1)=\frac{c_1^3+\barT(Q_1)\barZ(Q_1)}{1/a^2+Q_1}=\frac{c_1^3+\barT(Q_1)(\lambda^3/b^2)}{1/a^2+Q_1}.
\ee
This means
\be\label{TQ1_requirement}
\barT(Q_1)=\frac{(1/a^2+Q_1)a\lambda^2/b-c_1^3}{\lambda^3/b^2}
=(1/a^2+Q_1)\frac{ab}{\lambda}-\frac{b^2c_1^3}{\lambda^3}.
\ee
In addition, by the formula for $\barT$ in \eqref{barT_solution}, we have
\begin{align}
    \barT(Q_1)&=\frac{1/a^2+Q_1}{c_1} \int_{A_f}^{A_0} \frac{f(w)\,dw}{1+wf(w)^2}.
\end{align}
Thus, we must have
\begin{align}
    \int_{A_f}^{A_0} \frac{f(w)\,dw}{1+wf(w)^2}
    &=\frac{c_1 \barT(Q_1)}{1/a^2+Q_1}
    \\&=\frac{abc_1}{\lambda}-\frac{b^2 c_1^4}{\lambda^3 (1/a^2+Q_1)}
    \\&=\frac{b A_0}{a c_1}-\frac{1}{B_f}\label{c1_integral_eqn2}
\end{align}
where we have used the solutions for $\lambda$ and $Q_1$ in \eqref{lambda_formula} and \eqref{Q1_formula}.

Note that \eqref{c1_integral_eqn1} and \eqref{c1_integral_eqn2} exactly correspond to \eqref{first_integral_eqn}--\eqref{second_integral_eqn} with the association $\psi=c_1/b$. Thus, by Lemma~\ref{lemma:A0c1}, given $A_f,B_f>0$ where $A_f/B_f\le a^2$, there exists a unique pair $(A_0,c_1)$ satisfying both equations, where $A_0\ge A_f>0$ and $c_1>0$. Thus, from \eqref{lambda_formula}, we can conclude that $\lambda>0$. We can also use the formulas above to derive a set of solutions to the original differential equation system \eqref{DE1}--\eqref{DE_boundary}. We need to
verify that this solution satisfies the assumptions of Lemma~\ref{lemma:DE} in \eqref{DE_assumption}--\eqref{DE_bounding_assumption}. For \eqref{DE_assumption}, note that
\begin{align}
    (1+a^2S)(\lambda-R)+a^2T^2
    &=a^2\lambda^2(-\barS\barR+\barT^2)
    \\&=\frac{a^2\lambda^2\barS^2A}{c_1^2}
    \\&\ge \frac{\lambda^2A}{a^2c_1^2}
\end{align}
where the last step follows since $\barS\ge 1/a^2$. From the differential equations for $A$ and $B$ in \eqref{2d_system}, $A'\le 0$, which means $A(S)\ge A_f$ for all $S\in[0,Q_1]$. Thus,
\be
    (1+a^2S)(\lambda-R)+a^2T^2\ge \frac{\lambda^2 A_f}{a^2c_1^2}.
\ee
This quantity is indeed positive, proving \eqref{DE_assumption}.

Next we show \eqref{DE_bounding_assumption}, that all four functions are bounded. Certainly it is true for $A$, since $A_0\ge A(S)\ge A_f$ for $S\in[0,Q_1]$. Since $B(S)=f(A(S))$, and $f$ is continuous and bounded over any interval, $B$ is bounded. From \eqref{barT_solution}, we can conclude that $\barT$ is an increasing function, with $\barT(Q_1)$ bounded, which means $\barT(S)$ is bounded for all $S\in[0,Q_1]$. Given this, boundedness of $\barR,\barZ,\barV$ follow immediately from \eqref{barR_solution}--\eqref{barV_solution}. Finally, boundedness of $R,Z,V,T$ follow from the linear relationships in \eqref{bar_defs1}--\eqref{bar_defs2}.

It remains to confirm the quantities in the energy-per-bit expression. We already confirmed the formula for $Q_1$ in \eqref{Q1_formula} which is equivalent to the formula in the theorem \eqref{Q1_thm} with $\psi=c_1/b$. Using the expression for $\barT(Q_1)$ in \eqref{TQ1_requirement}, we may write
\begin{align}
    \frac{a}{b}T(Q_1)=\frac{a\lambda}{b}\barT(Q_1)&=1+a^2Q_1-\frac{abc_1^3}{\lambda^2}
    \\&=1+a^2Q_1-\frac{bA_0^2}{a^3c_1}.
\end{align}
Thus
\be
1+a^2Q_1-\frac{a}{b}T(Q_1)=\frac{bA_0^2}{a^3c_1}.
\ee
We also have
\begin{align}
    Z(0)&=\frac{\barZ(0)}{\lambda^2}-\frac{\lambda}{b^2}
    \\&=\frac{a^2 c_1^4 B_0}{\lambda^2}-\frac{\lambda}{b^2}
    \\&=\frac{A_0^2B_0}{a^2}-\frac{a^2c_1^2}{b^2 A_0}.
\end{align}
We have
\begin{align}
\barT(Q_1)&=(1/a^2+Q_1)\frac{ab}{\lambda}-\frac{b^2c_1^3}{\lambda^3}
\\&=\frac{b^3 A_0^4 B_f}{a^7c_1^4}-\frac{b^2A_0^3}{a^6 c_1^3}.
\end{align}
Thus
\be
T(Q_1)=\lambda \barT(Q_1)
=\frac{b^3 A_0^3 B_f}{a^5 c_1^2}-\frac{b^2 A_0^2}{a^4 c_1}.
\ee
Moreover
\begin{align}
    \barR(Q_1)&=\frac{\barT(Q_1)^2}{1/a^2+Q_1}-\frac{(1/a^2+Q_1)A_f}{c_1^2}
    \\&=\frac{a^6 c_1^2}{b^2 A_0^3 B_f}\left(\frac{b^3 A_0^4 B_f}{a^7c_1^4}-\frac{b^2A_0^3}{a^6 c_1^3}\right)^2-\frac{b^2 A_0^3 A_f B_f}{a^6 c_1^4}
    \\&=\frac{b^4 A_0^5 B_f}{a^8 c_1^6}-\frac{2 b^3 A_0^4}{a^7 c_1^5}+\frac{b^2 A_0^3}{a^6 c_1^4 B_f}-\frac{b^2 A_0^3 A_f B_f}{a^6 c_1^4}.
\end{align}
Thus
\begin{align}
R(Q_1)&=\lambda^2 \barR(Q_1)+\lambda
\\&=\frac{b^4 A_0^3 B_f}{a^4 c_1^2}-\frac{2 b^3 A_0^2}{a^3 c_1}+\frac{b^2 A_0}{a^2 B_f}-\frac{b^2 A_0 A_f B_f}{a^2}+\frac{a^2 c_1^2}{A_0}.
\end{align}
So we have
\be
\frac{aT(Q_1)}{b}-\frac{R(Q_1)}{b^2}
=-\frac{A_0}{a^2 B_f}+\frac{A_0 A_f B_f}{a^2}+\frac{b A_0^2}{a^3 c_1}-\frac{a^2 c_1^2}{b^2 A_0}.
\ee
We can now confirm the $Q_2$ value as
\be
\frac{aT(Q_1)}{b\lambda}-\frac{R(Q_1)}{b^2\lambda}
=-\frac{A_0^2}{a^4 c_1^2 B_f}+\frac{A_0^2 A_f B_f}{a^4 c_1^2}+\frac{b A_0^3}{a^5 c_1^3}-\frac{1}{b^2}.
\ee
Again using the association $\psi=c_1/b$ gives the formula in \eqref{Q2_thm}.
We can also confirm the value inside the log by combining several of the above to find
\begin{align}
    &1+Z(0)+a^2Q_1-\frac{2aT(Q_1)}{b}+\frac{R(Q_1)}{b^2}
    \\&=\left(1+a^2Q_1-\frac{aT(Q_1)}{b}\right)+Z(0)
    -\left(\frac{aT(Q_1)}{b}-\frac{R(Q_1)}{b^2}\right)
    \\&=\frac{A_0^2B_0}{a^2}
    +\frac{A_0}{a^2 B_f}-\frac{A_0 A_f B_f}{a^2}
    \\&=\frac{A_0}{a^2}\left(\frac{1}{B_f}+A_0B_0-A_fB_f\right).
\end{align}

\end{document}